\documentclass{article}

\usepackage{hyperref}

\usepackage{listings} 

\usepackage{graphicx}
\usepackage{amsmath}

\title{A derivation of the beam equation}

\author{Daniel Duque \\
  Canal de Experiencias Hidrodin\'amicas (CEHINAV) \\
  E.T.S. Ingenieros Navales\\
  Avda. Arco de la Victoria 4,
  28006 Madrid (Spain)}

\begin{document}

\maketitle

\begin{abstract}
  The Euler-Bernoulli equation describing the deflection of a beam is
  a vital tool in structural and mechanical engineering. However, its
  derivation usually entails a number of intermediate steps that may
  confuse engineering or science students at the beginnig of their
  undergraduate studies. We explain how this equation may be deduced,
  beginning with an approximate expression for the energy, from which
  the forces and finally the equation itself may be obtained. The
  description is begun at the level of small ``particles'', and the
  continuum level is taken later on. However, when a computational
  solution is sought, the description turns back to the discrete level
  again. We first consider the easier case of a string under
  tension, and then focus on the beam. Numerical solutions for several
  loads are obtained.
\end{abstract}

\section{Motivation}

The Euler-Bernoulli beam equation is of paramount importance in civil
engineering, being a simplified theory that yields relevant results
for the dynamics and statics of beams \cite{Gere,Timoshenko,beam}.
From the mathematical point of view, it is perhaps the simplest
differential equation of a high order (fourth) that has a clear
practical relevance.
%
However, a simple derivation is not easy to find, since most often a
number of intermediate concepts, such as bending moments and shear
forces, are introduced. We have designed a seminar in which the beam
equation is obtained from an expression for the energy.  As a
preliminary step, the equation for the string under tension is
derived, in a manner slightly different from what is usual. 
Computational methods are employed for both systems, in order to
find numerical solutions.

This article explains the method used. Its contents would fit in about
two hours, ideally followed by a practical computing session of about two
hours.

\section{Introduction}

Depending on the level of the class, some introduction to elementary
Newtonian particle mechanics may be needed. In particular, Newton's
Second Law, and the case when forces are conservative and can
therefore be obtained as derivatives of some potential energy.

However, Newton's laws deal with point particles and forces between
them. They are very accurate at scales such as the Solar
System, at which the planets and the Sun are almost point-like.
Another regime ln which Newtonian physics works very well is at
very small scales,
where molecules or atoms are again almost point-like (even if the
interactions themselves have a quantum origin).

Everyday objects, on the other hand, are usually \emph{extended}
(continuum). Historically, ``particles'' have been introduced in order
to apply point dynamics to these systems. Particles are portions of
material that are ``very'' small, yet still ``large enough'' to have
macroscopic features, such as density, volume, temperature, etc. In
practical terms, they can be as small as small cells, about $1$ $\mu$m
in size. If they are smaller, thermal effects can cause such effects
as Brownian motion. Of course, the students should not relate these
particles, with real particles, such as electrons.

The idea behind this approach is to apply Newtonian physics to each of
these particles.  Then, one takes the \emph{continuum limit}, in which
the number of these particles, $N$, is taken to infinity, but
some of their features are taken zero (others may stay tend to infinity
or reach a finite value.) As a simple example, the
mass of each particle, $m$ could tend to zero in such a way that the
total mass, given by
$$
M=Nm
$$
remains constant. The same applies for lengths and other quantities,
even if in some cases, as we will see, the correct limit is not
obvious to anticipate.

This way, one may derive laws expressed as partial differential
equations.  For example, we will derive here the wave equation:
\[
\frac{\partial^2 y}{\partial t^2} = v^2 \frac{\partial^2 y}{\partial
  x^2}
\]
and the beam equation:
\[
\frac{\partial^2 y}{\partial t^2} = -EI \frac{\partial^4 y}{\partial
  x^4} + q
\]

Traditionally, many special mathematical methods have been devised to
solve these equations, leading to huge advances in mathematics,
physics, and engineering \cite{arfken}. The advent of computing has changed
the situation somewhat, often providing a more direct and
easier way to find solutions (or more precisely, a numerical
approximation to the solutions.)  On the other hand computers are
\emph{discrete} by nature. A continuum equation can therefore not be
computed as is: it must be discretized.  The conclusion, as we will
discuss, is that we return to where we started, two centuries ago
\cite{patankar}.

\section{The string}

This is system covered in many textbooks, but we treat it here for two
reasons. The first is that the procedure will be mirrored when
discussing the beam, but the expressions are simpler in this case.
The second is that some of the quantities that are introduced in
elementary textbooks (chiefly, the tension) arise naturally in this
procedure.

We model a string as a line of point particles (we will call ``beads'')
joined by massless springs. Of course, most strings are not built this
way, but the point is that the continuum limit will be the same for a
wide number of systems, and we choose a simple one to work with.

\subsection{Energy}

The expression for the energy of a spring comes from Hooke's law:
\[
U= \frac12 \kappa (\ell - \ell_0)^2 ,
\]
where $\ell_0$ is the spring's natural length and $\ell$, its actual
length. The elastic parameter $\kappa$ is Hooke's spring
constant. This is the best known expression, but here we will prefer
to use an expression that features the relative deformation, $(\ell -
\ell_0)/\ell_0$, called the ``strain'':
\[
U = \frac12 \frac{B}{\ell_0} (\ell - \ell_0)^2 ,
\]
where $B=\kappa \ell_0$ is a parameter with units of force, whose
meaning will be discussed later.

\begin{figure}
  \begin{center}
    \includegraphics[width=0.7\textwidth]{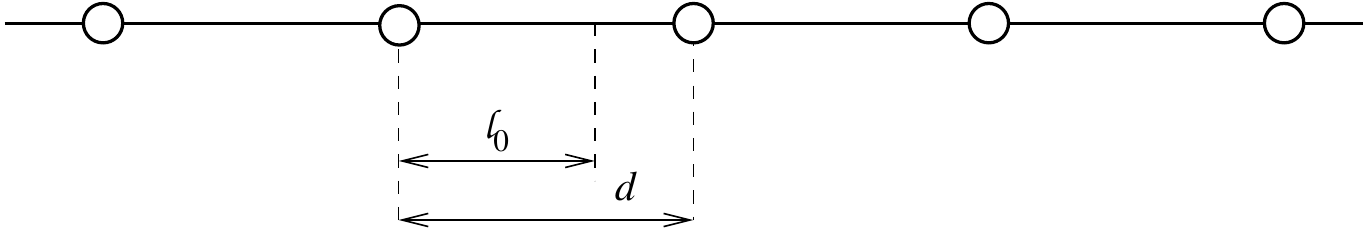} \\
    \includegraphics[width=0.7\textwidth]{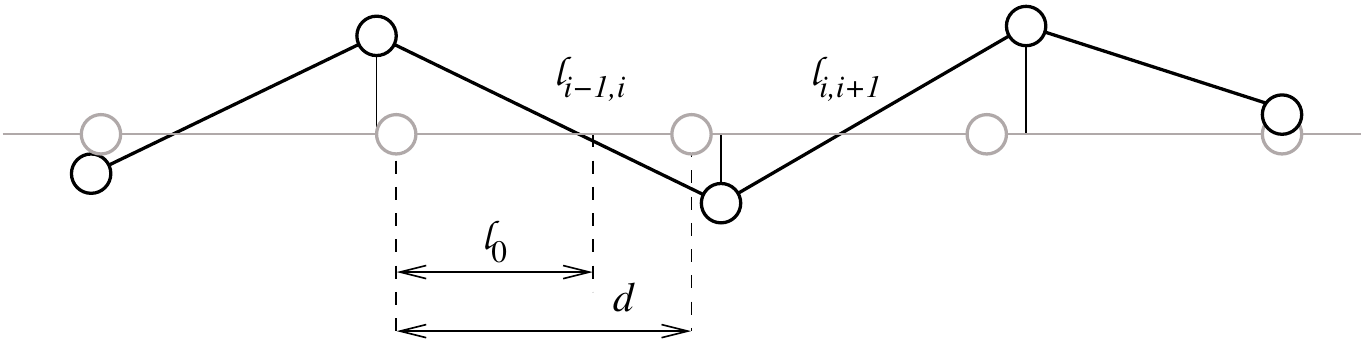}
  \end{center}
  \caption{\label{fig:string1}
    The string under tension, unperturbed (upper graph),
    and perturbed (lower graph) }
\end{figure}

Let us first consider the string under tension, but unperturbed
otherwise.  Its shape will be a straight line as in the upper part of
Figure \ref{fig:string1}, and all the spring lengths will be equal to
$d$, with $d>\ell_0$. The energy will therefore be
\[
U_0= N \times \frac12 \frac{B}{\ell_0} ( d - \ell_0) ^2 = 
\frac12 \frac{B}{\ell_0 N } ( d N - \ell_0 N) ^2 =
\frac12 \frac{B}{L_0} ( L - L_0) ^2 .
\]

In order to reach the later equality we have multiplied and divided by
$N$, and we have written $Nd=L$ for the length of the string and
$N\ell_0 = L_0$ for its natural length.  This is actually our first
application of the continuum limit, since we begin with an expression
that depends on the particles' magnitudes $d$ and $\ell_0$ and we end
up with another one that depend on the macroscopic magnitudes $L$ and
$L_0$. Also, the parameter $B$ seems to require no change with $N$ in
order the energy $U$ be finite. Since $B=\kappa \ell_0$, the string
constant $\kappa$ should then tend to infinity (not to zero!) in this
limit.

If, on the other hand, the string is distorted, as in the lower part
of Figure \ref{fig:string1}, the energy will now be:
\[
U= \frac12 \frac{B}{\ell_0} (\ell_{i-1,i} - \ell_0)^2 +
   \frac12 \frac{B}{\ell_0} (\ell_{i,i+1}-\ell_0)^2  +\cdots  ,
\]
where, out of the $N$ terms, we just write the two of them that are
related to the $i$-th bead at the center of Figure \ref{fig:string1}.
We will use $i$ subindices to refer to beads and $i-1,i$ to refer to
quantities between bead $i-1$ and bead $i$, such as the length of the
spring between them (similarly for $i,i+1$).

\subsection{Forces and tension}

Let us find first the horizontal component of the force on bead $i$:
\[
f_{i}^x = - \frac{\partial U}{\partial x_i} =  
- \frac{B}{\ell_0} (\ell_{i-1,i} - \ell_0) \frac{\partial\ell_{i-1,i}}{\partial x_i}
- \frac{B}{\ell_0} (\ell_{i,i+1} - \ell_0) \frac{\partial\ell_{i,i+1}}{\partial x_i} ,
\]
where he have applied the chain rule of differentiation to $\ell_{i-1}$
and $\ell_{i+1}$. The Pythagorean theorem tells us:
\[
\ell_{i-1,i}^2= (x_i-x_{i-1})^2 + (y_i-y_{i-1})^2  \qquad   \ell_{i,i+1}^2= (x_{i+1}-x_i)^2 + (y_{i+1}-y_i)^2 .
\]
From this, we can obtain these interesting identities:
\[
\frac{\partial\ell_{i-1,i}}{\partial x_i} =   \cos \theta_{i-1,i} \qquad  
\frac{\partial\ell_{i,i+1}}{\partial x_i} = - \cos \theta_{i,i+1} 
\]

Therefore:
\[
f_{i}^x =    - \frac{B}{\ell_0} \left[
  (\ell_{i-1,i} - \ell_0) \cos\theta_{i-1,i} -
  (\ell_{i,i+1} - \ell_0) \cos\theta_{i,i+1} 
  \right] .
\]

We can write this as
\[
f_{i}^x =    - T_{i-1,i}  \cos\theta_{i-1,i} +  T_{i,i+1}  \cos\theta_{i,i+1} ,
\]
where the tension between bead $i-1$ and bead $i$ being
$T_{i-1,i}=\frac{B}{\ell_0} (\ell_{i-1,i} - \ell_0)$, similarly for $T_{i+1,i}$.

Now, if deflections are small: $\ell_{i-1,i}, \ell_{i,i+1} \approx d $, therefore
$T_{i-1,i}=T_{i,i+1}=T$, where
\[
T=  B\frac{d - \ell_0 }{\ell_0} = B\frac{L - L_0 }{L_0} 
\]
Moreover, $\cos\theta_{i,i+1}\approx 1$ for all beads, therefore $f_x\approx 0$.

Notice this tension $T$ is the external force to be applied to the
ends of the string to keep it tout. Indeed, the left end particle
($i=1$) has no neighbor at its left to pull from it, and the tension
$T$ must be applied from the outside. The same applies to the other
end.  Moreover, recalling the total energy is $U_0= \frac12
\frac{B}{L_0} ( L - L_0) ^2 $, we see $T ' = - dU /dL = -T$.  What
this means is that the string is trying to shrink with a force $T'$
that we must overcome with another one $T$ which is equal but pointing
outwards.

The $B$ parameter is then seen to be
\[
B =   T\frac{L_0}{L-L_0 } =    \frac T{(L-L_0)/L_0 } .
\]
One of the important elastic properties of a material is its Young's
modulus (also known as tensile modulus, or elastic modulus), a
magnitude with units of pressure that is defined as the stress/strain
ratio:
\[
E =  \frac{T/A_0}{(L-L_0)/L_0 } .
\]
The strain is, as in Hooke's law, $(L-L_0)/L_0$, and the stress is the
tension divided by the cross section of the string under no tension,
$A_0$.  Therefore, our $B$ parameter is related to Young's modulus:
$$B=E A_0.$$
As a simple experiment, students can try to measure experimentally
values of Young's modulus from these equations, see Appendix
\ref{sec:experiments}.

The vertical component of the force follows from the identity
\[
\frac{\partial\ell_{i-1,i}}{\partial y_i} =   \sin \theta_{i-1,i} \qquad
\frac{\partial\ell_{i,i+1}}{\partial y_i} = - \sin \theta_{i,i+1} ,
\]
with the end result:
\[
f_{i}^y =    - T_{i-1,i}  \sin\theta_{i-1,i} +  T_{i,i+1}  \sin\theta_{i,i+1} ,
\]

Many physics books, such as \cite{alonso,Tipler,Sears}, basically
start with this equation for the vertical force, from considerations
of the net vertical on each bead. Our derivation has the advantage of
providing more insight on the meaning of the tension $T$. It is also
less likely to result in errors in signs. On the other hand, in
\cite{Bauer} we find a derivation similar to ours (although it is
given in terms of rods coupled by torsion.)

In the limit of small vertical deflections, the force may be written
as
\[
f_{i}^y=    - T  \sin\theta_{i-1,i} +  T \sin\theta_{i,i+1} 
\]

In this limit, the sines are also similar to the tangents, so:
\[
f_{i}^y \approx   - \frac{T}{d} ( y_i - y_{i-1} ) +  \frac{T}{d} (y_{i+1} - y_i) =
\frac{T}{d} (y_{i-1} -2 y_i + y_{i+1} )
\]

\subsection{The wave equation}

Let us continue with the equations of motion for our bead.  Newton's
Second Law gives (only the $y$ direction is changing, so we will drop
the $y$ superindices):
\[
m a_i=   \frac{T}{d} (y_{i-1} -2 y_i + y_{i+1} )
\]
that can be written as
\[
a_i= \frac T{m/d}  \frac{ y_{i-1}   -2 y_i + y_{i+1} }{d^2}  =
\frac T{\mu}  \frac{ y_{i-1} -2 y_i + y_{i+1}   }{d^2}  ,
\]
with $\mu = m / d$ the mass per unit length.

The last ratio is a discrete, finite differences, version of the
second spatial derivative \cite{numerical}.  Therefore, in the limit
$d\rightarrow 0$ we may write the wave equation
\[
a=\frac{\partial^2 y}{\partial t^2} = \frac T\mu \frac{\partial^2
  y}{\partial x^2} .
\]
It can be shown that the phase velocity of traveling waves is given by
$v^2=T/\mu$.

\subsection{The loaded string}

It is often interesting to find the equilibrium solution to equations,
setting the time derivatives equal to zero.
In this case, it is rather dull: 
the solution to $\frac{\partial^2 y}{\partial x^2} =0$ is just a straight line.
For homogeneous Dirichlet boundary conditions $y(0)=y(L)=0$, the unique
solution is simply $y(x)=0$.

To make things more interesting, we may add a vertical force $F(x)$ to
each bead. This force is constant in the vertical direction, but may
vary along the string. The energy equation would be modified as:
\[
U = \cdots - F(x) y . 
\]

The wave equation is now:
\[
\frac{\partial^2 y}{\partial t^2} = \frac T\mu \frac{\partial^2
  y}{\partial x^2} + F/m 
\]

The static solution is given by the equation:
\[
\frac T\mu \frac{\partial^2 y}{\partial x^2}  = -F/m 
\]

This is a Poisson equation. For example, for the case of gravity one
would have
\[
F=-mg \qquad\rightarrow\qquad \frac T\mu \frac{\partial^2 y}{\partial x^2}  = g .
\]

Setting $y(0)=y(L)=0$, the unique solution is a parabola:
\begin{equation}
\label{eq:parabola}
y= -  \frac{\mu g }{2 T} x (L-x) .
\end{equation}

Some students may know the solution to this sort of problems involving
hanging strings is often more involved, with shapes such as the
catenary resulting. That is the case, but in this limit of small
deformations the solution is simply an upward parabola, which is the usual
limit of any curve close to a minimum.

\subsection{Computing the loaded string}

A computer may be used in order to find the equilibrium shape of a
string under general loads. However, in order to apply computational
methods we need to go back to the discretized equations, which are the
ones that are readily implemented on a computer. This actually takes
us back to the historic derivation of these equations, as we have
discussed.

For example, we would have the equation of motion:
\[
a_i= \frac T\mu  \frac{ y_{i-1}   -2 y_i + y_{i+1} }{d^2}  + F_i/m .
\]
For the static case:
\[
\frac{ y_{i-1} -2 y_i  + y_{i+1}  }{d^2}  = - \frac{F_i}{Td} = q_i,
\]
where $q_i=-F_i/(Td)$ is the string load.

Notice this is a linear equation that cannot readily be solved only for
$y_i$, since it involves $y_{i-1}$ and $y_{i+1}$, which are also unknown.  In
fact, one has a system of $N$ linear equations, which be written in
matrix form:
\[
\underbrace{%
\frac1{d^2}
\begin{pmatrix}
  \ddots & \vdots & \vdots & \vdots & \vdots & \vdots & \\
  \cdots & 1 & -2 & 1 & 0 & 0  & \cdots \\
  \cdots & 0 & 1 & -2  & 1 & 0  & \cdots  \\
  \cdots & 0 & 0 &  1 & -2 & 1 & \cdots \\
  & \vdots & \vdots & \vdots & \vdots & \vdots & \ddots \\
\end{pmatrix}
}_{\nabla^2}
\underbrace{%
\begin{pmatrix}
  \vdots \\
  y_{i-1} \\
  y_i    \\
  y_{i+1} \\
  \vdots \\
\end{pmatrix}
}_{\vec{y}} 
=
\underbrace{%
\begin{pmatrix}
  \vdots \\
  q_{i-1} \\
  q_i      \\
  q_{i+1} \\
  \vdots \\
\end{pmatrix}
}_{\vec{q}} .
\]

It can also be summarized as the symbols under the under-braces suggest:
\[
\nabla^2 \vec{y} = \vec{q} ,
\]
where $\nabla^2$ is a matrix for second derivatives, $\vec{y}$ is a
vector containing the vertical positions, and $\vec{q}$ is a vector
containing the loads.

This linear algebra problem is implemented in all major computational
environments. We choose to carry out the calculations using
\lstinline|python|, it being a powerful emerging ``scientific
ecosystem'' with many advantages \cite{scipy}. One of them is that it
is free and open source, and is included in all major
\lstinline|linux| distributions. Another choice with the same
advantages is \lstinline|octave|, which is designed to be a clone of
\lstinline|matlab|, itself a viable choice but not free. Other options
such as \lstinline|maple| or \lstinline|Mathematica| are also
possible.

We recommend \lstinline|ipython| in notebook form, to open an
interactive session within a web browser in the lecture room, and run
the programs ``live''. Relevant files can be obtained in the
Supplementary Materials section of this article.

\begin{figure}
  \begin{center}
    \includegraphics[width=0.7\textwidth]{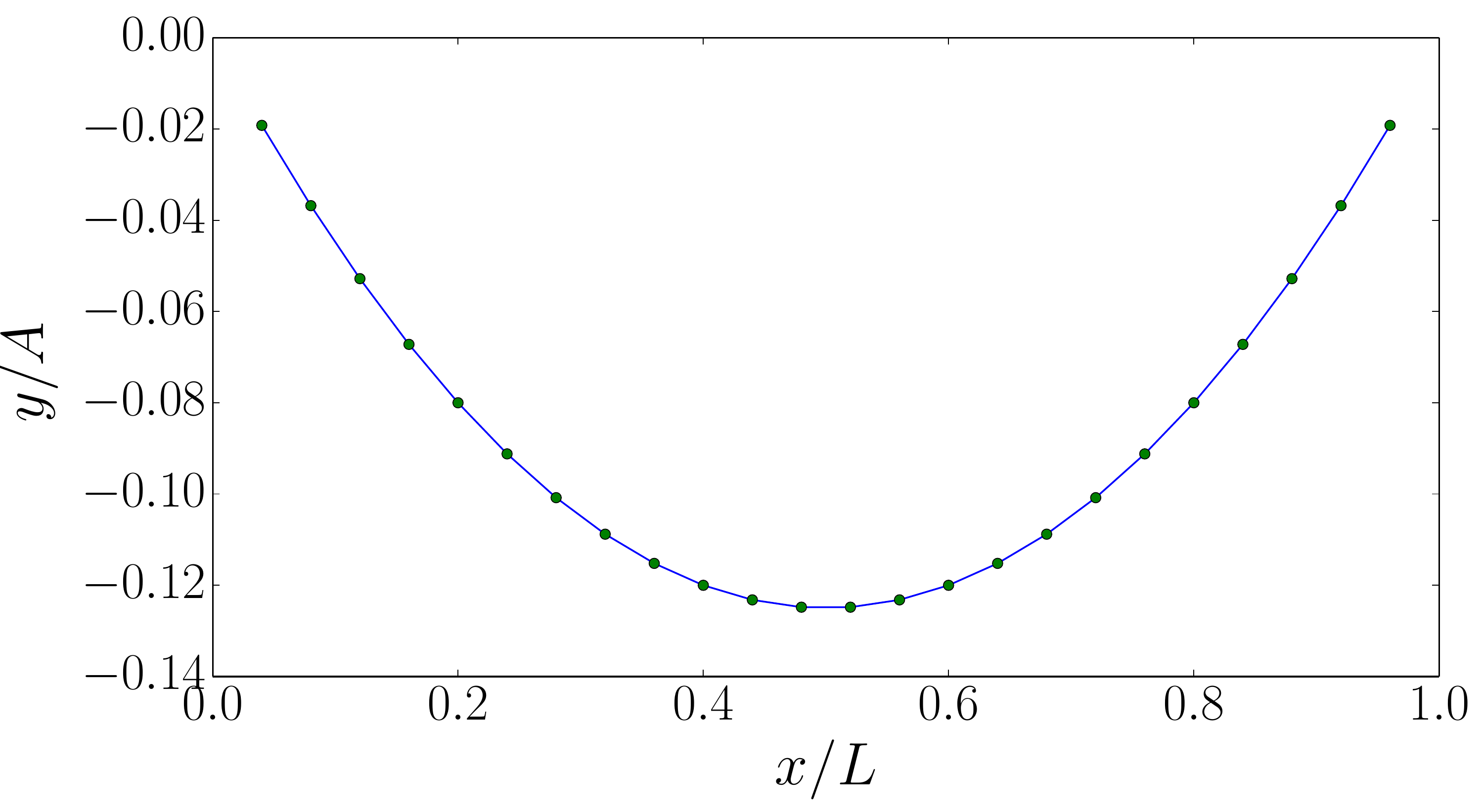} \\
    \includegraphics[width=0.7\textwidth]{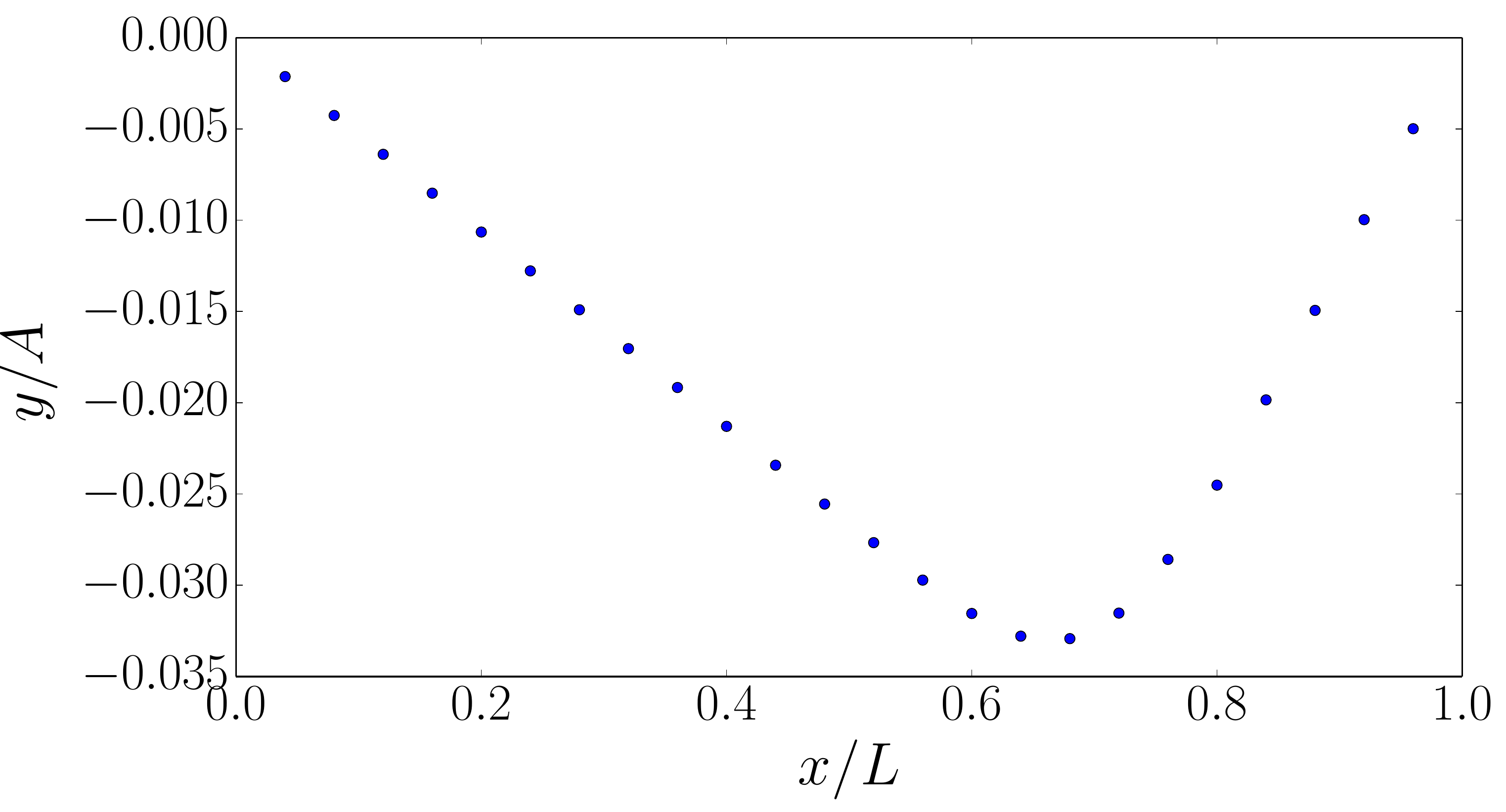} 
  \end{center}
  \caption{\label{fig_string_1} Numerical result for the uniformly
    loaded string (above) and for a complex loading (below). Dots:
    numerical results, line: theoretical solution.  Horizontal length
    scaled by $L$, and vertical displacement by $A:=\frac{\mu g L^2 }{
      T }$.}
\end{figure}

In Figure \ref{fig_string_1} (above) we plot the solution for the
shape of the equilibrium string under uniform load.  The agreement
with the exact parabolic solution of (\ref{eq:parabola}) is seen to be
excellent. Indeed, this approximation to the second derivative is
known to be exact (up to machine precision) for quadratic functions.

Some students may notice that, for certain choice of parameters, the
numerical $y$ values may not be ``small'' and can be actually
comparable to $L$. However, they should be reminded that variables are
usually cast into non-dimensional form in computations. For example,
writing (\ref{eq:parabola}) in the form:
\[
y= -  \frac{\mu g L^2 }{2 T} \frac x L \left(1-\frac x L \right) ,
\]
we see it is not $x$ but rather $x/L$ which is relevant, and that the
scale of $y$ is fixed by $\frac{\mu g L^2 }{T}$. It is the later
length scale, and how it compares with $L$, which tells us whether our
string has little deformation or not.

It would seem not so useful to obtain a solution that is already
known, but this numerical method still works for loads that are not so
simple.  As an example, we show results for a Gaussian load $ q(x)=-
\exp (- ( (x-0.7) / 0.1)^2) $ (in reduced units), which which may
model, e.g. some deformation due to a blunt object. In Figure
\ref{fig_string_1} (below) we show that the solution has a shape that
could be expected, with linear parts on the zones where little load is
applied.

\section{The beam}

\begin{figure}
\begin{center}
  \includegraphics[width=0.6\textwidth]{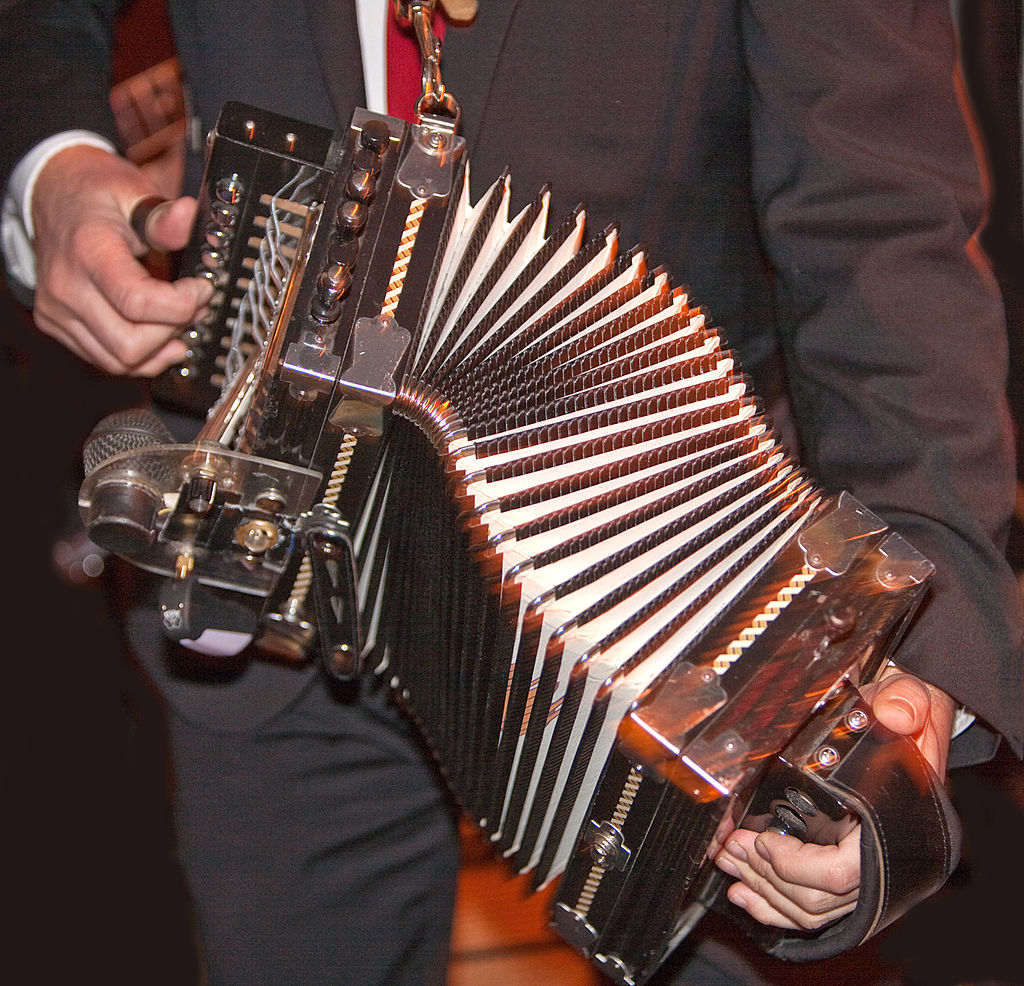} \\
\end{center}
\caption{\label{fig:accordion}
  An accordion, a mental image of our model for a beam \cite{accordion}
}
\end{figure}

In the beam we are concerned with \emph{bending}, not compression or
expansion. We may picture a physical beam as a succession of slabs that
are subject to bending. It would be like a accordion, an instrument
featuring a bellows that resembles a secession of slabs, see Figure
\ref{fig:accordion}. It would be a silent one, since the instrument
emits sounds when compressing or expanding the bellows,
as air enters or leaves it, but not when it is bent.

\begin{figure}
\begin{center}
  \includegraphics[width=0.6\textwidth]{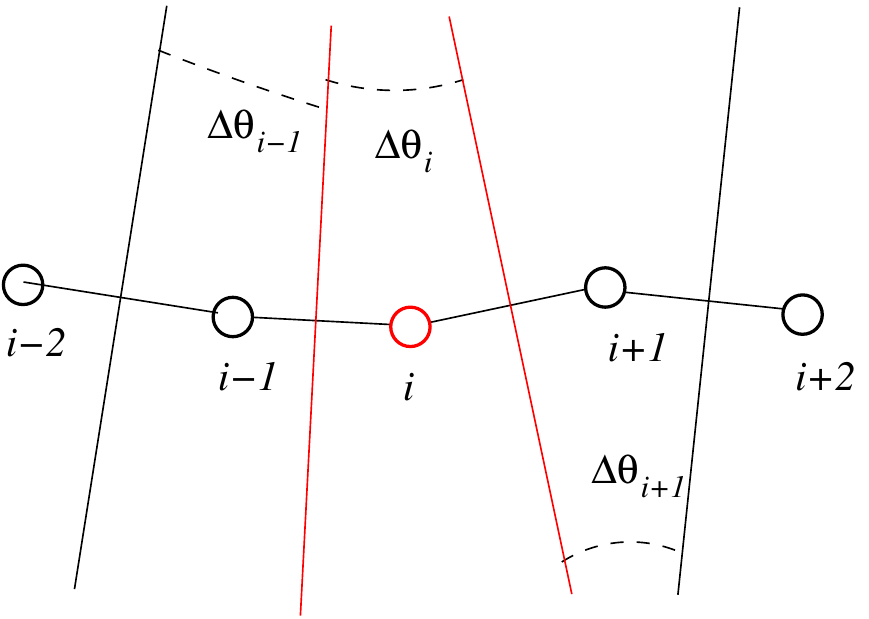}
\end{center}
\caption{\label{fig:beam1}
A model of a beam as a series of slabs}
\end{figure}

\subsection{Energy and forces}

We begin by writing the energy as:
\[
U=\frac12 \frac{C}{d}
\left[ 
  (\Delta \theta_{i-1})^2 +        (\Delta \theta_i)^2 +       (\Delta \theta_{i+1})^2  +\cdots 
\right]
\]
where $C$ is a stiffness parameter with units of force $\times$
area. Each $\Delta\theta_i$ is the difference of angles limiting each
slab, see Figure \ref{fig:beam1}. What we are supposing here is that
each slab has its energy increased when its two limiting angles
differ, and that this dependence is quadratic to lowest order (it
cannot be linear since a reversal in sign of the angles should lead to
the same energy increase.)

\begin{figure}
\begin{center}
  \includegraphics[width=0.4\textwidth]{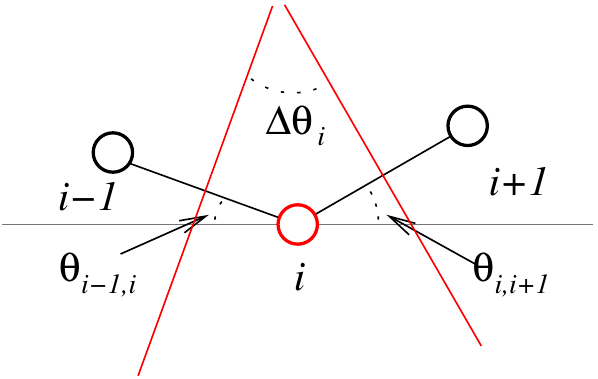}
\end{center}
\caption{\label{fig:beam2}
  Detail of a single slab}
\end{figure}

As we can see in Figure \ref{fig:beam2}
\[
\Delta\theta_i= \theta_{i,i+1} -    \theta_{i-1,i}  \approx
\frac{y_{i+1}-y_i}{d} -  \frac{y_i-y_{i-1}}{d} =
\frac{ y_{i+1}   -  2 y_i + y_{i-1}  }{d} .
\]
We therefore find again the discrete version of the second derivative
(but for a $1/d$ factor. )

The energy can now be written as
\[
U=\frac12 \frac{C}{d^3}
\left[ 
  (2 y_{i-1} - y_i    - y_{i-2} )^2 +
  (2 y_i    - y_{i+1} - y_{i-1} )^2 +
  (2 y_{i+1} - y_{i+2} - y_i    )^2  +\cdots 
\right]
\]

It is not too difficult to obtain the $y$ component of the force on slab
$i$:
\begin{equation}
  \label{eq:4force}
  f_{i}^y = - \frac{\partial U}{\partial y_i} =   \frac{C}{d^3}
  \left[ 
    y_{i-2}  
    - 4y_{i-1} +  6 y_i 
    - 4 y_{i+1}
 +  y_{i+2}
  \right]  
\end{equation}

This is an approximation to the fourth derivative (but for a $
-1/d$ factor ) \cite{numerical}. Therefore:
\[
f(x) \approx -  C d \frac{\partial^4 y}{\partial x^4}
\]

\subsection{The beam equation}

It is again easy to add vertical forces to each slab:
\[
U = \cdots - F(x) y 
\]

The final dynamical equation is
\[
m  \frac{\partial^2 y}{\partial t^2} = -  C d  \frac{\partial^4 y}{\partial x^4}  + F  ,
\]
and its static solution is given by:
\[
C \frac{\partial^4 y}{\partial x^4}  =     \frac Fd = q 
\]
where $q=F / d$ is the beam load.  As shown in the Appendix
\ref{sec:appendix}, the $C$ parameter is $C=EI$, where $E$ is Young's
modulus (again) and $I$ a quantity known as the second moment of
inertia. We therefore obtain the dynamic beam equation:
\[
\mu  \frac{\partial^2 y}{\partial t^2} =   - EI   \frac{\partial^4 y}{\partial x^4}  +  q 
\]
and its static version, which is probably better known:
\[
E I \frac{\partial^4 y}{\partial x^4} = q
\]

\subsection{Computing the loaded beam}

Again, our discrete problem involving the slabs can be cast as a linear
algebra problem:
\[
\nabla^4 \vec{y} = \vec{q} ,
\]
with a fourth derivative matrix $\nabla^4$ having: a diagonal with
$-6$ values, subdiagonals above and below it with values of $4$, and
finally subdiagonals above and below the two former ones with values
of $-1$, with a common factor of $1/d^4$, as seen in Equation
(\ref{eq:4force}).

\begin{figure}
\begin{center}
  \includegraphics[width=0.7\textwidth]{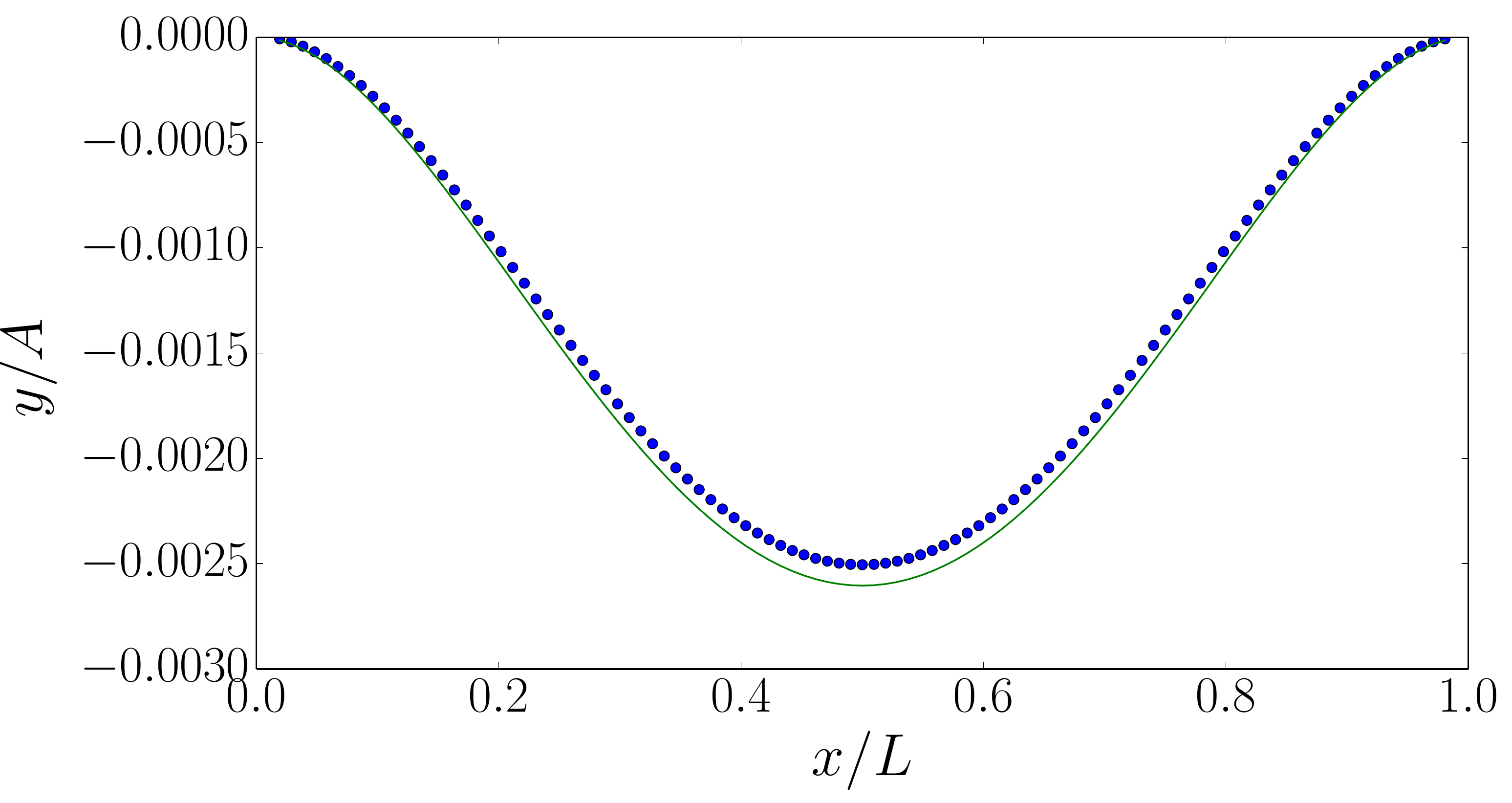} \\
  \includegraphics[width=0.7\textwidth]{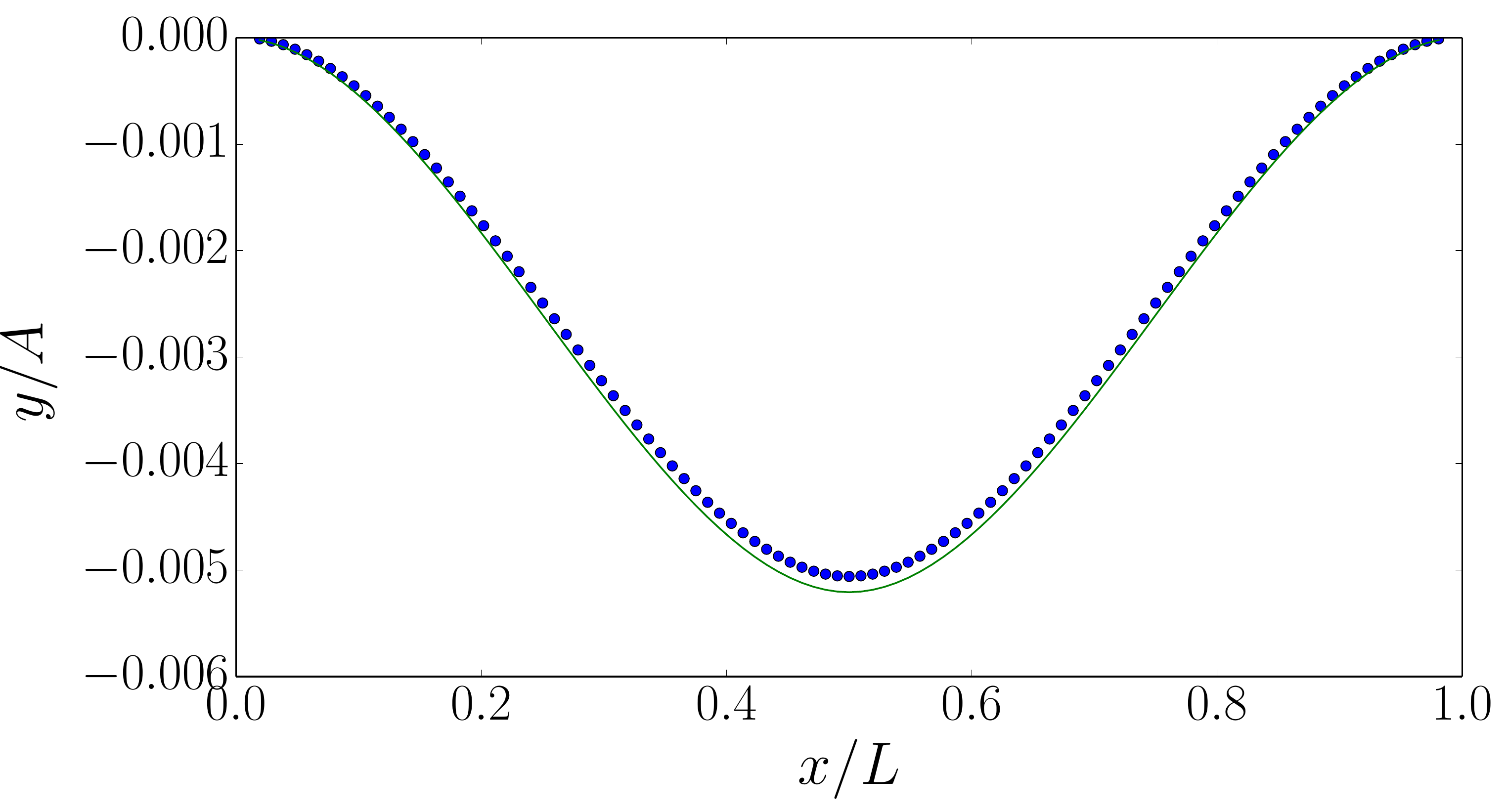} \\
  \includegraphics[width=0.7\textwidth]{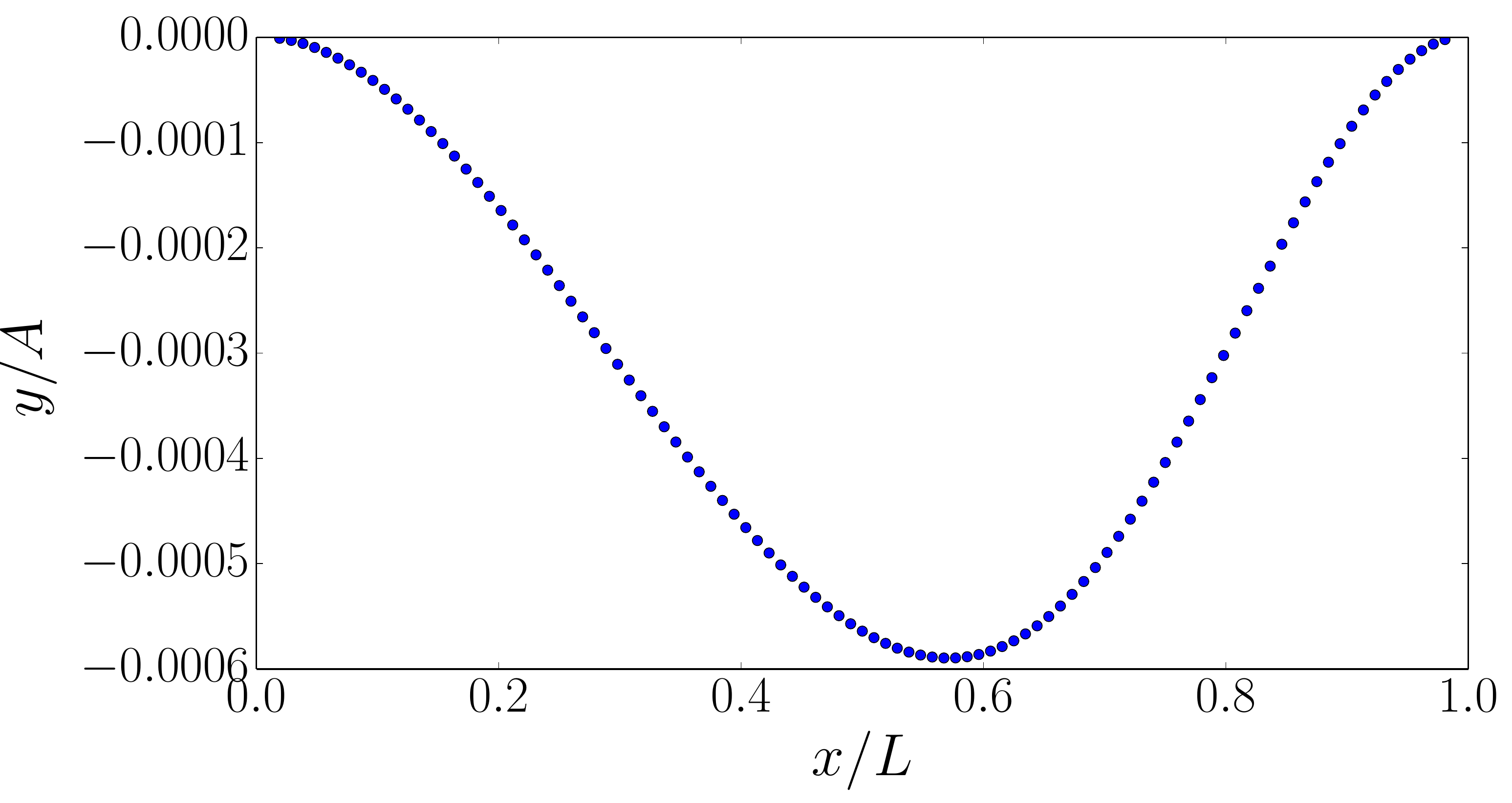} \\
\end{center}
\caption{\label{fig:beam_1} Numerical result for the uniformly loaded
  string (above), centrally loaded beam (middle) and for a complex
  loading (below). Dots: numerical results, lines: theoretical
  solutions. Horizontal length scaled by $L$, and vertical
  displacement by $A:=\frac{q_0 L^4 }{E I}$.}
\end{figure}

In Supplementary Materials the code to solve the static loaded beam
can be found. We consider here a double clamped beam, in which the
ends are fixed to be $0$ and both ends, and so are the first
derivatives. The result for uniform loading, Figure \ref{fig:beam_1}
(above), is not so accurate when compared with the exact solution
\cite{beam} as the string was, since this is a higher order derivative
for which the approximation is not so good. Nevertheless, the
numerical solution can be seen to converge to the exact one as $N$ is
increased.

We may also consider the case of central loading, where the whole
load is placed in the middle of the beam. The resulting beam
shape is plotted in Figure \ref{fig:beam_1} (middle), and compared with the
exact solution. Finally, we reuse our
Gaussian load function for the string and apply it to the beam. Since
a beam is different from a string, the resulting shape, Figure
\ref{fig:beam_1} (below) is not so obvious to guess, with the maximum
deformation away from $x = 0.7 L$, the point at which the load is
greater.  The students are encouraged to perform simple experiments, as
described in Appendix \ref{sec:experiments}.

\section{Conclusions}

We have shown in this lecture how the concept of ``particle'' may be
used in order to obtain physical laws written as differential
equations. A traditional point of view is to take these laws as the
ultimate expressions, for which solutions should be obtained in
different situations. Mathematically this entails that a given
differential equation has different solutions corresponding to
different boundary conditions and initial conditions. However, in
later years the emergence of computers makes it easy to
obtain numerical solutions to the equations. Since computers are
discrete, the equations must be brought into discrete form, which
actually brings us back to particles.

This situation may seem paradoxical, but most experienced researchers
will agree that computational techniques do not replace, but rather
compliment, traditional mathematical analysis. However, the direct
simulation of a particle description can, in our opinion, be a
powerful teaching resource for first year college courses.

There are many ways in which this lecture may be extended. Additional
simple experiments may be proposed in addition to the ones
given at the Appendix \ref{sec:experiments}.

\section*{Acknowledgments}

This article comes from a lecture given at Kyoto University for
students of the International Degree in Civil
Engineering. Profs. H. Gotoh and A. Khayyer and are warmly thanked for
the invitation that made this possible.  Also, thanks to the students,
whose feedback during and after the lecture has greatly improved the
presentation of this material. Funding from the Spanish Ministry for
Science and Innovation under grant TRA2013-41096-P ``Optimization of
liquid gas transport for LNG vessels by fluid structure interaction
studies'' is acknowledged.

\section*{References}

\bibliographystyle{unsrt}
\bibliography{beam_bib}

\appendix

\section{The stiffness parameter}
\label{sec:appendix}

A slab is compressed in different amount at different values of
$y$. Indeed the compression (or extension) at height $y$ is:
\[
c=y \Delta\theta \qquad y\in(-h/2,h/2) .
\]
In the later range for $y$ we are already assuming deviations are
small.

\begin{figure}
\begin{center}
  \includegraphics[width=0.3\textwidth]{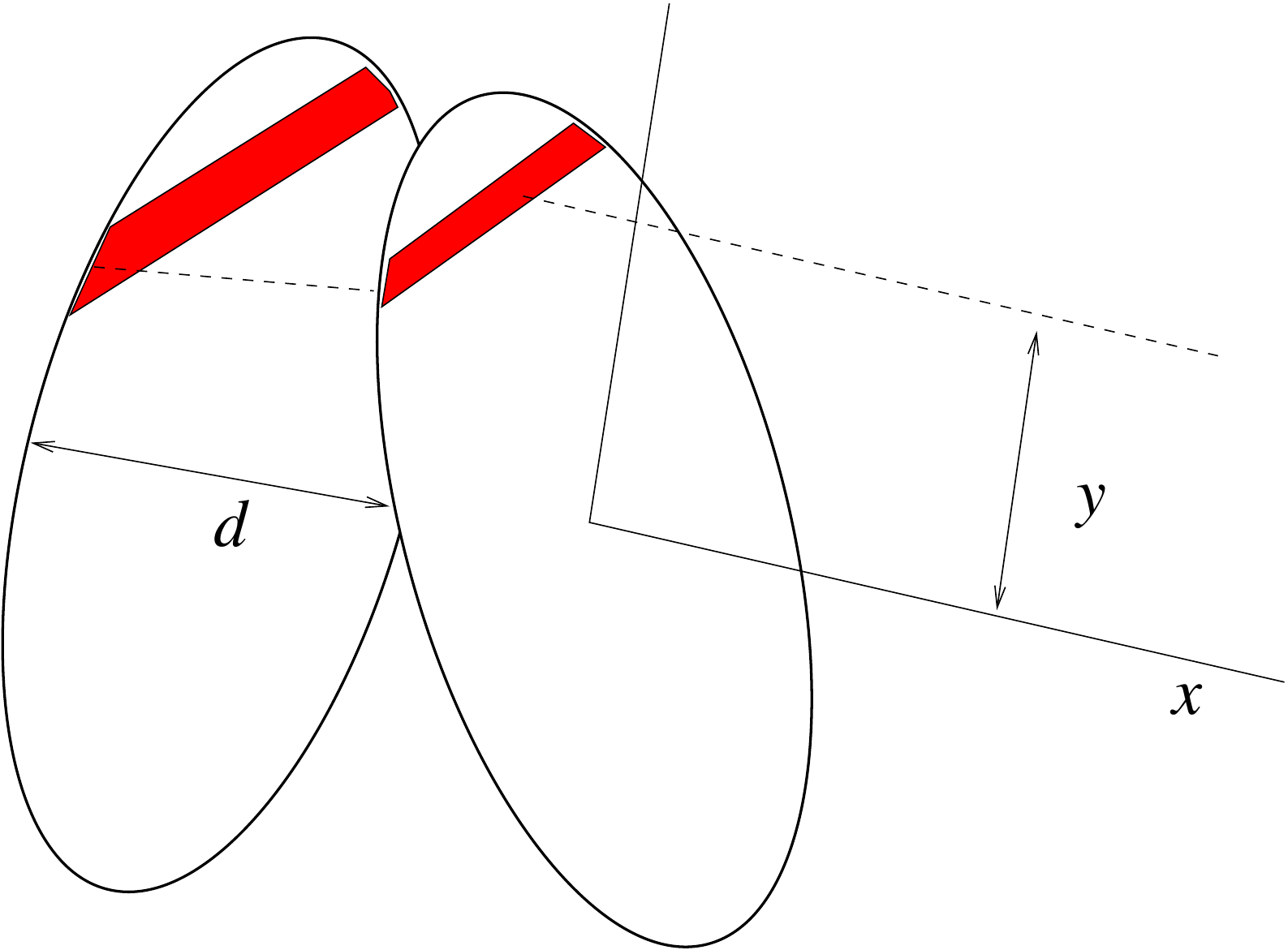}
\end{center}
\caption{Calculation of the compression energy for a slab \label{fig:beam3}}
\end{figure}

The energy cost of a compression for a strip with area $dA$ (see
Figure \ref{fig:beam3}) is:
\[
dU=\frac12 (E dA) \frac{c^2}{d} ,
\]
where $E$ is Young's modulus.

The total energy cost for this slab is therefore
\[
U_0=\frac12 \frac{E}{d} \int_{-h/2}^{h/2} c^2 dA = \frac12 \frac{E}{d}
(\Delta\theta)^2 \left(\int_{-h/2}^{h/2} y^2 dA \right)
\]

The last integral is purely geometric, and is called $I$ ``the second
moment of area'' (aka moment of inertia of plane area, area moment of
inertia, or second area moment), with units of area squared. Therefore
\[
U_0= \frac12 \frac{E I }{d} (\Delta\theta)^2
\]

Recall our original expression for each of the slabs:
\[
U_0=\frac12 \frac{C}{d}   (\Delta \theta)^2  .
\]

Clearly, $C$ is given by $C=EI$.

\section{Experiments}
\label{sec:experiments}

As a simple experiment, students can try to measure experimentally
values of Young's modulus from our equations $B=E A_0$ and $T= B (L -
L_0 )/L_0 $. This means we can obtain $B$ and $E$ by measuring the
length of the string under no tension, its length when tuned, and its
tension. Ideally, all of these quantities should be measured, but some
of them can be taken from the manufacturer. For example, a standard
$.036$ in guitar string means a diameter of $0.9144$ mm that may be
not easy to measure. This experiment can also be combined with
elementary wave theory. If the string is tuned to a known fundamental
frequency $f$, then $ f=\frac1{2L}\sqrt{\frac T\mu} $, a fact that may
let us measure $T$ given $L$ and $\mu$.

A simple experiment for the beam is to clamp the end of a flexible
object, such as a ruler, and measure its deflection at the hanging
end. This would be a uniformly-loaded cantilever beam if the load
results simply from the weigth of the object. Other loadings can
of course be explored. The experiments
are quite easy to carry out, but the correct mathematical and numerical
escription of the hanging end needs to be carefully addressed. If the object
vibrates, there is also a relationship for the frequency, which is
more complicated than for the string, but whose solution can be
found in standard books \cite{beam}.

\end{document}